\documentclass[showpacs,amsmath,amssymb, prb,twocolumn]{revtex4}
\usepackage{graphicx}
\usepackage{dcolumn}
\usepackage{float}
\usepackage{bm}
\begin{document}

\title{Cooper pairs localization in tree-like networks of superconducting islands}

\author{Francesco Romeo$^{\ast}$}
\affiliation{ Dipartimento di Fisica ''E. R. Caianiello'', Universit\`a degli Studi di Salerno,
Via Giovanni Paolo II, I-84084 Fisciano (Sa), Italy\\
$^{\ast}$email: fromeo@sa.infn.it (corresponding author)\\
\texttt{http://orcid.org/0000-0001-6322-7374}}
\author{Roberto De Luca}
\affiliation{ Dipartimento di Fisica ''E. R. Caianiello'',
	Universit\`a degli
	Studi di Salerno, Via Giovanni Paolo II, I-84084 Fisciano (Sa), Italy}

\begin{abstract}
We study inhomogeneous Cooper pairs distribution and localization effects in tree-like networks of superconducting islands coupled via Josephson weak links. Using a generalized Feynman's approach, reminiscent of the Bose-Hubbard model, we demonstrate that the Cooper pairs fraction which localizes on a specific network's island is limited by the network topology and, if present, by the repulsive interaction. These findings contribute to clarify the interplay between confinement effects induced by the network's topology and interaction and shed some light on recent experiments dealing with networks of Josephson junctions.
\end{abstract}
\maketitle
\section{Introduction}
\label{sec:intro}
The discovery of superconductivity\cite{onnes} and its theoretical explanation\cite{BCS1,BCS2}, i.e. the Bardeen-Cooper-Schrieffer (BCS) theory, are among the most important achievements in physics. The BCS paradigm successfully explains a large variety of phenomena ranging from the organization of nuclear matter\cite{nuclearBook,nulcearBCS} to the superconductivity in metals\cite{degennes} and conjectures exist about the possibility that the final quantum state of a black hole would be similar to the ground state of a BCS superconductor\cite{BH1,BH2}.\\
Apart from its cultural centrality, superconductivity is nowadays an essential ingredient to implement topological states of matter\cite{majorana} (e.g., the Majorana's modes) and quantum computers\cite{qcomputer}. For this reason, there is a renewed interest towards the possibility of conditioning the superconductivity by using different means. Traditionally, new superconducting materials have been identified by doping a parent compound in a controlled fashion. Even though the outcome of this procedure is hardly predictable, in fortunate cases it provides superconducting alloys with enhanced critical temperature. Extensive application of this procedure led to the discovery of high critical temperature (HTC) superconductors\cite{HTC} and the impact of this success has fueled the dream of finding a room-temperature superconductor for a long time.\\
Recently, it has been proposed that dielectric metamaterials can be used to create effective media in which the electron-electron effective interaction (depending on the inverse of the dielectric function) can be enhanced\cite{enz1,enz2}. The enhancement of the interaction between electrons requires the ability of engineering the dielectric function of an artificially designed system. In this way a near-zero dielectric function (ENZ) metamaterial with enhanced superconducting critical temperature can be obtained. The aforementioned theoretical framework shows a good agreement with the experimental realizations\cite{enz3,enz4} (e.g. Al-Al$_2$O$_3$ core-shell metamaterials) and predicts the possibility to achieve near room-temperature superconducting systems based on conventional superconductivity\cite{roomBCS}.\\
Interestingly, the enhancement of the superconducting critical temperature of a metamaterial based on a conventional superconductor with bulk gap $\Delta$ (and coherence length $\xi_{BCS} \propto \Delta^{-1}$) implies that the resulting superconducting gap $\Delta_{eff}$ is also enhanced compared to $\Delta$. Provided that the BCS relation between superconducting gap and coherence length is also valid for the effective medium, the above observation suggests that the coherence length of the effective medium $\xi_{eff}$ is shortened compared to $\xi_{BCS}$. This conclusion is not surprising since the typical distance $\sim \xi_{eff} \propto \Delta_{eff}^{-1}$ between the paired electrons in a Cooper pair depends on the effective interaction strength $g_{eff} \propto \Delta_{eff}$.\\
A less obvious implication of this analysis is that superconducting gap and critical temperature of a synthetic superconductor can be enhanced by reducing the effective coherence length of the system.\\
A reduced coherence length is observed in superconductors altered by the introduction of an appropriate amount of non-magnetic disorder (impurities). These systems are described under appropriate circumstances \cite{Randeria2001} by the dirty limit coherence length $\xi_{eff} \approx \sqrt{\ell \ \xi_{BCS}}$ [\onlinecite{notaPippard}], which on its turn depends both on the coherence length $\xi_{BCS}$ of the clean system and the disorder-dependent mean free path $\ell$. Assuming that $\Delta_{eff} \propto \xi_{eff}^{-1}$, one easily get $\Delta_{eff} \propto \Delta \sqrt{\xi_{BCS}/\ell}$. Thus, a gap enhancement (i.e., $\Delta_{eff}>\Delta$) would be expected when the effective coherence length $\xi_{eff}$ is limited by the mean free path (i.e. when $\ell < \xi_{BCS}$). The condition $\ell< \xi_{BCS}$ is hardly reachable by using HTC superconductors because they are characterized by very short coherence length (few nanometers in cuprates\cite{xiCuprate}). Thus, it is expected that the disorder-induced gap enhancement works well for low critical temperature superconductors.\\
These arguments are not in contradiction with the Anderson's theorem [\onlinecite{ATh}] according to which nonmagnetic disorder does not affect the superconducting transition temperature in conventional superconductors. The Andreson's statement, indeed, is formulated under the restrictive condition of space-independent order parameter, which typically is not realized in disordered systems. A recent generalization of the Anderson's argument [\onlinecite{gATh}] confirms the possibility of enhancing the superconductivity by using disorder. Thus the above arguments appear to be plausible provided that $\Delta_{eff}$ is interpreted, e.g., as space-averaged order parameter.\\
Despite the aforementioned conclusions are based on rather naive arguments, their validity can be proven to some extent by using a microscopic theory based on the self-consistent solution of the Bogoliubov-de Gennes equations\cite{gastiasoro18}. In particular, a critical temperature enhancement has been reported for disordered superconducting systems in close vicinity of the Anderson localization condition\cite{burmistrov12,burmistrov15,multifractality}. Thus, exponential localization seems to be a favorable condition to obtain a critical temperature enhancement.\\
Exponential localization of the superconducting order parameter can be also obtained in tree-like networks of superconducting islands coupled via Josephson weak links. These systems, originally proposed as analogue model to study Bose-Einstein condensation on inhomogeneous graphs\cite{burioni,burioni2001,buonsante}, have been the object of intense and systematic experimental investigations\cite{silvestrini,lorenzo,ottaviani,lucci}. Experiments confirm that network nodes with higher connectivity act as localization centres for the Cooper pairs. Moreover, indirect signatures of a gap enhancement are also reported\cite{lucciBis}. The above evidences suggest that these artificial systems mimic the interstitial structure of the superconducting order parameter established at the phase boundary between the superconducting state and an insulating state (see Fig. 4(n) of Ref. [\onlinecite{gastiasoro18}]). This intuition is corroborated by a de Gennes-Alexander micronetworks formulation\cite{microNetBook} of the problem showing a close connection between order parameter localization and critical temperature enhancement\cite{romeoGL}. In these systems, exponential localization of the order parameter is controlled by the network's topology (instead of the impurity potentials) and for this reason the graph connectivity directly affects the superconducting phase transition \cite{nota00}. In view of the relevance of the Cooper pairs localization phenomenon in graph-like structures, it would be important to verify the existence of mechanisms limiting the particles accumulation on a given system's island.\\
In this work we demonstrate that the Cooper pairs fraction which localizes on a specific network's island is limited by the network topology and, if present, by the repulsive interaction. Treating Cooper pairs like ordinary bosons, we derive a generalized Feynman's model through which inhomogeneous Cooper pairs distribution and localization effects are systematically studied.\\
We will focus on tree-like structures which appear to be an ideal platform to study localization phenomena induced by the network topology.
Differently from arbitrary graphs, two nodes (vertices) of a tree-like network are connected by a unique path, so that interference effects have a limited influence on the particles wavefunction. Moreover, the absence of loops limits unintentional inductive couplings and thus it is a desirable condition for the experimental implementation using Josephson junctions' networks.\\
The work is organized as follows. In Sec. \ref{sec:fey} we review the Feynman's model of a Josephson junction. Its generalization obtained by the Bose-Hubbard model is presented in Sec. \ref{sec:bose-hub}. The resulting model is a non-linear Schr\"{o}dinger equation whose ground state solution allows to study the interplay between repulsive interaction and network connectivity. Selected applications of the theory are discussed both in the presence (Sec. \ref{sec:int}) or in the absence (Sec. \ref{sec:nullint}) of repulsive interaction. Conclusions are given in Sec. \ref{sec:concl}, while technical details are presented in Appendix \ref{app:A}, \ref{app:C} and \ref{app:B}.

\section{Feynman's model of a Josephson junction}
\label{sec:fey}
In his famous lecture \cite{feynman}, Feynman proposed a simple model of Josephson junction\cite{josephson,baroneBook, delucaBook}. According to the model, the dynamics of the macroscopic wavefunctions of the superconducting electrodes, i.e. $\psi_1$ and $\psi_2$, is described by the equations:
\begin{eqnarray}
\label{eq: fey}
i\hbar \ \dot{\psi}_1=E_1 \psi_1-K\psi_2\nonumber\\
i\hbar \ \dot{\psi}_2=E_2 \psi_2-K\psi_1,
\end{eqnarray}
where the dot notation indicates the time derivative. The constant $K$ describes the weak coupling of the two sides of the junction and depends on the characteristics of the tunnel barrier. The presence of an external circuit is implicitly assumed. It induces a difference between the ground state energies of the two condensates, so that $E_1-E_2=qV$ represents such difference written in terms of the absolute vale of the Cooper pair charge $q=2e$ and the voltage drop across the junction $V$. The macroscopic wavefunctions of the Cooper pairs condensates can be written in the form $\psi_\alpha=\sqrt{\rho_{\alpha}}\exp(i \theta_{\alpha})$ ($\alpha \in \{1,2\}$), where $\rho_{\alpha}$ and $\theta_{\alpha}$ represent the Cooper pairs density and the phase factor pertaining to the electrode labelled by $\alpha$, respectively. Using the exponential representation of the macroscopic wavefunctions in Eq.(\ref{eq: fey}) and introducing the phase difference $\varphi=\theta_2-\theta_1$, we get:
\begin{eqnarray}
\label{eq:fey1}
\dot{\rho}_2=-\dot{\rho}_1=\frac{2K}{\hbar}\sqrt{\rho_1 \rho_2} \sin(\varphi)\\
\label{eq:fey2}
\dot{\theta}_1=-\frac{U_1}{\hbar}+\frac{K}{\hbar}\sqrt{\frac{\rho_2}{\rho_1}}\cos(\varphi)\\
\label{eq:fey3}
\dot{\theta}_2=-\frac{U_2}{\hbar}+\frac{K}{\hbar}\sqrt{\frac{\rho_1}{\rho_2}}\cos(\varphi).
\end{eqnarray}
The dynamics of the phase difference $\varphi$ is easily deduced by using Equations (\ref{eq:fey2})-(\ref{eq:fey3}) and can be written as follows:
\begin{eqnarray}
\label{eq:phase}
\dot{\varphi}=\dot{\theta}_2-\dot{\theta}_1=\frac{qV}{\hbar}+\frac{K}{\hbar}\Bigl (\sqrt{\frac{\rho_1}{\rho_2}}- \sqrt{\frac{\rho_2}{\rho_1}}\Bigl ) \cos(\varphi).
\end{eqnarray}
Due to the action of the external circuit, tunneling phenomena across the junction do not change the Cooper pairs densities $\rho_1$ and $\rho_2$, so that one can assume $\rho_1 = \rho_2 = \rho_0$, being $\rho_0$ the time-independent particles density of the isolated electrodes. The implementation of the latter conditions is problematic within the Feynman's model and, to be consistent, one has to assume $\rho_1=\rho_0-\delta \rho$ and $\rho_2=\rho_0+\delta \rho$ with $\delta \rho/\rho_0 \ll 1$. Under these assumptions and after a first-order expansion in $\delta \rho$, Eq. (\ref{eq:fey1}) is written as:
\begin{eqnarray}
\label{eq:j1}
\delta \dot{\rho}\simeq \frac{2K}{\hbar}\rho_0 \sin(\varphi)\equiv J_0 \sin(\varphi),
\end{eqnarray}
which represents the correct current-phase relation of the Josephson junction. Under the same assumptions, Eq.(\ref{eq:phase}) takes the form:
\begin{eqnarray}
\label{eq:phaseApprox}
\dot{\varphi}\simeq\frac{qV}{\hbar}-\frac{2K}{\hbar}\Big (\frac{\delta \rho}{\rho_0}\Big) \cos(\varphi),
\end{eqnarray}
the latter showing a deviation from the strict voltage-frequency relation:
\begin{eqnarray}
\label{eq:j2}
\dot{\varphi}=\frac{qV}{\hbar}.
\end{eqnarray}
The deviation, consisting in a cosine term in the right-hand side of Eq. (\ref{eq:phaseApprox}), originates from the simplifying approximations of the model. Indeed, in real systems the external circuit, neglected within the Feynman's approach, maintains $\delta \rho=0$. The latter constraint can be implemented within the semiclassical treatment proposed by Ohta\cite{ohta}, where the condition $\dot{\rho}_1=\dot{\rho}_2=0$ is exactly respected.\\
Despite these limitations, the Feynman's model provides the correct constitutive equations of the Josephson effect (Equations (\ref{eq:j1}) and (\ref{eq:j2})) when a zero-order expansion in $\delta\rho$ is considered. From the physics viewpoint, the latter expansion is justified, e.g., when the coupling constant $K$ is so small that the Cooper pairs densities remain practically unchanged ($\rho_{1,2} \approx \rho_0$) as long as the time $\tau \ll \hbar/(2 K)$ is reached.\\
We will show that the cosine term in Eq. (\ref{eq:phaseApprox}), which is not expected in the voltage-frequency relation of a biased Josephson junction, is a necessary ingredient to get an appropriate description of isolated systems (i.e., not connected to external reservoirs). Thus, in the following, we present a generalization of the Feynman's model allowing the description of isolated networks of superconducting islands coupled via the Josephson effect.

\section{From Bose-Hubbard to the generalized Feynman's model}
\label{sec:bose-hub}
We are interested in studying the Cooper pairs distribution in tree-like networks of superconducting islands. Cooper pairs are composite particles obeying hard-core bosonic algebra, being the deformed bosonic algebra reminiscent of the fermionic nature of the paired electrons. Treating the Cooper pairs as genuine bosons, which is certainly an approximate picture, we can study the tree-like systems mentioned above by means of the Bose-Hubbard model. This approximate picture is not too crude. For instance, quantum critical phenomena in granular superconductors have been investigated by modeling the Cooper pairs as a quantum fluid of charged bosons [\onlinecite{fisher1988}]. Similarly, quantum phase transitions of Josephson junctions arrays are often described by adopting a short-range Bose-Hubbard model [\onlinecite{fisher1989,bruder92,burioni,rizzi2006}], being the latter equivalent, under appropriate conditions, to the second quantization Hamiltonian of a Josephson junctions array [\onlinecite{fisher1989}]. Second quantization Hamiltonian of the Bose-Hubbard model describing a tree-like system is written as
\begin{eqnarray}
\label{eq:bh}
H=\sum_{i}\epsilon_i n_i-K \sum_{ij}\mathcal{A}_{ij}b^{\dag}_i b_j+\frac{U}{2}\sum_{i}n_i(n_i-1),
\end{eqnarray}
where $\epsilon_i$ is a site-dependent energy, $K>0$ is a coupling constant, $U>0$ is the strength of the repulsive interaction and $n_i=b^{\dag}_i b_i$ is the number operator of the lattice site $i$ written in terms of bosonic creation and annihilation operators. The islands connectivity is defined by the adjacency matrix of elements $\mathcal{A}_{ij}$. Adjacency matrix is real and symmetric with vanishing diagonal elements ($\mathcal{A}_{ii}=0$). Islands labelled by the site index $i$ and $j$ are linked if $\mathcal{A}_{ij}=1$, while they are disconnected if $\mathcal{A}_{ij}=0$. Using Heisenberg equation of motion and bosonic commutation relations $[b_i,b_j]=[b^{\dag}_i,b^{\dag}_j]=0$ and $[b_i,b^{\dag}_j]=\delta_{ij}$ we get:
\begin{eqnarray}
\label{eq:eom}
i \hbar \frac{d}{dt}b_k=\epsilon_k b_k-K\sum_{j}\mathcal{A}_{kj}b_j+U n_k b_k.
\end{eqnarray}
Under macroscopic occupation of the islands (i.e. $\langle b^{\dag}_kb_k\rangle=N_k\gg 1$), the annihilation operator $b_k$ in Eq. (\ref{eq:eom}) can be substituted with the condensate wavefunction $\psi_k=\sqrt{N_k}\exp(i \theta_k)$ [\onlinecite{brunelli}]. Moreover, due to the latter assumption, fluctuations are suppressed and the condensate wavefunction depends on the well-defined semiclassical variables $N_k$ and $\theta_k$ [\onlinecite{smerzi}]. In this way, the discrete non-linear Schr\"{o}dinger equation\cite{NLSAtoms}
\begin{eqnarray}
\label{eq:nse}
i \hbar \frac{d}{dt}\psi_k=\epsilon_k \psi_k-K\sum_{j}\mathcal{A}_{kj}\psi_j+U |\psi_k|^2 \psi_k
\end{eqnarray}
is obtained. Equation (\ref{eq:nse}) preserves the total number of particles and thus the macroscopic wavefunction of the condensate can be normalized to the total number of Cooper pairs $N_T$ according to the relation $\sum_{k}|\psi_k|^2=N_T$. Interestingly, Feynman's model given in Eq. (\ref{eq: fey}) can be obtained from Eq. (\ref{eq:nse}) by considering a two-sites system with vanishing repulsive interaction ($U=0$). For these reasons, Eq. (\ref{eq:nse}) can be considered a generalization of the Feynman's model. In the following analysis, we will use Eq. (\ref{eq:nse}) to characterize the inhomogeneous distribution of Cooper pairs in tree-like networks of identical superconducting islands. The nonlinear term in Eq. (\ref{eq:nse}) mimics, at mean-field level, the interaction between Cooper pairs \cite{nota0}. The latter contribution is expected to be negligible in bulk superconductors, while it may be relevant when superconducting islands of reduced volume are considered. In view of our interest in isolated systems with identical superconducting islands, we can also set $\epsilon_k=0$ in Eq. (\ref{eq:nse}). Disorder effects (not treated in this work) can be studied by considering random site's potentials and randomly distributed hopping terms.\\

\subsection{Searching for the ground state of the system: general procedure}
\label{sec:NLSnum}
The ground state of the system can be obtained by considering the ansatz $\psi_k=\mathcal{C}_k \exp(i \theta)$, with $\mathcal{C}_k=\sqrt{N_k}$, which ensures vanishing Josephson currents between the superconducting islands (see Appendix \ref{app:A}). When the trial wavefunction is substituted in Eq. (\ref{eq:nse}), the following relations are obtained:
\begin{eqnarray}
\label{eq:ans1}
&& \dot{\mathcal{C}}_k=0\\
\label{eq:ans2}
&& -\hbar \dot{\theta} \ \mathcal{C}_k=-K \sum_{j}\mathcal{A}_{kj}\mathcal{C}_j+U \ \mathcal{C}^3_k.
\end{eqnarray}
A close inspection to Eqs. (\ref{eq:ans1})-(\ref{eq:ans2}) shows that $\dot{\theta}$ cannot depend on time and thus we set $\dot{\theta}=\omega$, with $\omega$ a time independent constant. The identification of the ground state requires the solution of a non-linear eigenvalues problem (see Eq. (\ref{eq:ans2})) with eigenvalues of the form $\mu=-\hbar \omega$.\\
Before treating the general case, it is quite instructive considering the $U=0$ case for which we have to solve an ordinary eigenvalues problem. In this case, the problem is described by a pure hopping Hamiltonian $H=-K \mathcal{A}$ which, in real space, is a real and symmetric matrix proportional to the adjacency matrix $\mathcal{A}$ via the constant $-K$. Thus, the hopping Hamiltonian and the adjacency matrix are simultaneously diagonalizable and share a common set of eigenvectors. The eigenproblem of the adjacency matrix is written as $\mathcal{A} \ \mathcal{C}=\lambda \ \mathcal{C}$ with $\lambda=\hbar \ \omega/K=-\mu/K$. The eigenvector $\mathcal{C}_M$ of the adjacency matrix with greatest eigenvalue $\lambda_M$ also corresponds to the unique ground state of the system in view of the relation $H\mathcal{C}_M=\mu_0 \mathcal{C}_M $ with eigenvalue $\mu_0=-K \lambda_M$. The spectral graph theory (Perron-Frobenius theorem) prescribes that the components of $\mathcal{C}_M$ are all positive, while $\lambda_M$ is bounded from above by the maximum degree of the graph [\onlinecite{nota1}]. Moreover, there are no other positive eigenvectors of the adjacency matrix (except positive multiples of $\mathcal{C}_M$) so that the theorem also implies the unicity of the bosonic ground state. Once $\mathcal{C}_M$ is known, it provides information about the particles distribution. These arguments show the close connection between graph theory and the ground state properties of the bosonic system.\\
The inclusion of a finite repulsive interaction modifies the aforementioned picture and the nonlinear problem has to be solved numerically. Numerical procedure is conveniently implemented by introducing eigenvectors normalized to $1$ instead of the total number of Cooper pairs $N_T$. Thus, we set $\mathcal{C}_k=\sqrt{N_T} \mathcal{B}_k$ with $\sum_k \mathcal{B}^2_k=1$ in Eq. (\ref{eq:ans2}). Using the above notation, the nonlinear problem takes the following form:
\begin{eqnarray}
\label{eq:nlp}
\lambda \ \mathcal{B}_k= \sum_{j}\mathcal{A}_{kj}\mathcal{B}_j-\xi \ \mathcal{B}^3_k
\end{eqnarray}
with $\xi=N_T U/K$ a small, but finite, perturbation of the linear problem. As before, the greatest eigenvalue of the nonlinear problem in Eq. (\ref{eq:nlp}) is associated to the ground state eigenvector. The solution of the nonlinear problem in Eq. (\ref{eq:nlp}) can be obtained by using an iterative procedure which is similar to that customarily used when a many-body system is studied within the Hartree-Fock approximation. The initial nonlinear problem is mapped into the linear eigenproblem\cite{salerno}
\begin{eqnarray}
\label{eq:nlpMap}
\lambda \ \mathcal{B}_k= \sum_{j}(\mathcal{A}_{kj}+V_{kj})\mathcal{B}_j
\end{eqnarray}
complemented by the self-consistency condition $V_{kj}=-\xi \delta_{kj}\mathcal{B}^2_j$. The solution is iteratively determined according to the following steps. The matrix $V_{kj}$ is computed by using the eigenstate with the greatest eigenvalue (EGE) of the noninteracting problem ($\xi=0$). Thus, eigenvectors and eigenvalues of $\mathcal{A}+V$ are determined. The EGE is used to recompute $\mathcal{A}+V$. After diagonalization of the latter operator, a new estimate of the EGE is obtained and the procedure can be iterated. The algorithm ends when the self-consistency condition $|V_{kj}+\xi \delta_{kj}\mathcal{B}^2_j|<\varepsilon$ is reached with accuracy $\varepsilon$. The procedure typically presents fast convergence with accuracy very close to the machine precision ($ \sim \ 10^{-12}$ in dimensionless units). Once the convergence has been reached, the solution of the nonlinear problem in Eq. (\ref{eq:nlp}) is obtained and the spatial distribution of the particles can be determined.\\

\begin{figure}[h]
\includegraphics[scale=0.25]{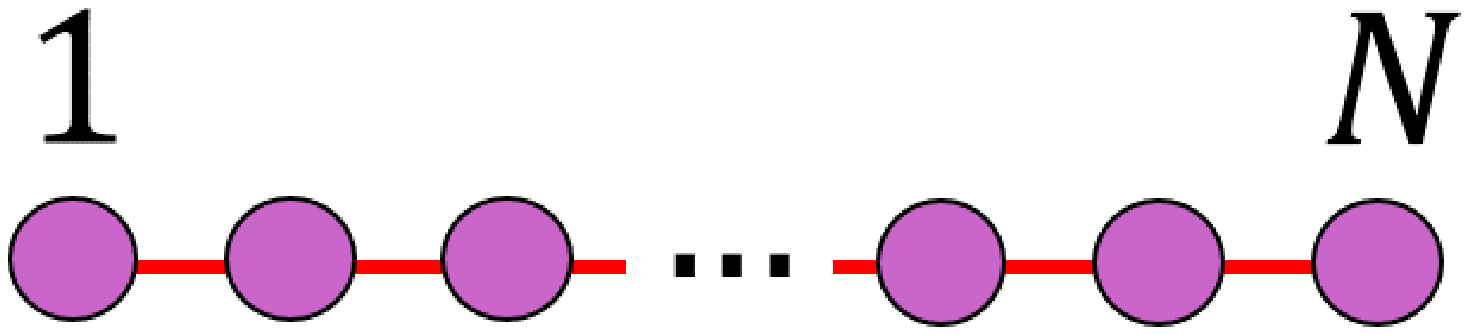}
\caption{Linear chain of islands. The straight lines represent the connections between islands indicated by circles. The degree of a terminal island is given by $\mathbb{D}_1=\mathbb{D}_N=1$, while the degree of any other node is given by $\mathbb{D}_j=2$ ($j \neq 1,N$).}
\label{figure1}
\end{figure}

\section{Selected applications of the theory: non-interacting case ($U=0$)}
\label{sec:nullint}
In this section, we present selected applications of the theory for which analytical solutions are available. In this way, the particles distribution can be studied avoiding numerical diagonalization of the adjacency matrix. In view of their pedagogical value, fully worked examples are reported.
\subsection{Linear chain}
In the following we will explain in detail the problem of the Cooper pairs distribution on a linear chain (Fig. \ref{figure1}). In particular, we consider $N_T$ non-interacting bosonic particles (i.e. the Cooper pairs) constrained to move on a linear connection of $N$ superconducting islands. Within this assumptions, the time evolution of the bosonic wavefunctions is described by a pure hopping Hamiltonian leading to the following equation:
\begin{eqnarray}
\label{eq:linear}
i\hbar \partial_t \psi_n=-K (\psi_{n-1}+\psi_{n+1}).
\end{eqnarray}
Using $\psi_n=\sqrt{N_n}\exp(i \theta_n)$ in Eq. (\ref{eq:linear}), we get the coupled equations:
\begin{widetext}
\begin{eqnarray}
\dot{N}_n &=& -\frac{2K}{\hbar}\Big [\sqrt{N_{n-1}N_{n}}\sin(\theta_{n-1}-\theta_{n})+\sqrt{N_{n+1}N_{n}}\sin(\theta_{n+1}-\theta_{n})\Big]\\
\dot{\theta}_n &=& \frac{K}{\hbar}\Big[\sqrt{\frac{N_{n-1}}{N_n}} \cos(\theta_{n-1}-\theta_{n})+\sqrt{\frac{N_{n+1}}{N_n}} \cos(\theta_{n+1}-\theta_{n})\Big],
\end{eqnarray}
\end{widetext}
showing the presence of cosine terms which, as anticipated in Sec. \ref{sec:fey}, are crucial for a correct description of the isolated system. When a stationary state of the closed system is considered, the number of particles residing on a single island does not change in time ($\dot{N}_n=0$). This requirement can be obtained by setting $\theta_n=\theta(t)$, with $\dot{\theta}(t)=\omega$ a time-independent constant. Accordingly, using the notation $\lambda=\hbar \omega/K$ and $\mathcal{C}_n=\sqrt{N_n}$, we obtain the difference equation:
\begin{eqnarray}
\label{eq:difflin}
\mathcal{C}_{n-1}+\mathcal{C}_{n+1}=\lambda \ \mathcal{C}_n,
\end{eqnarray}
complemented by the boundary conditions $\mathcal{C}_{0}=\mathcal{C}_{N+1}=0$. The difference equation in Eq. (\ref{eq:difflin}) can be presented in the form $\mathcal{A}\mathcal{C}=\lambda \mathcal{C}$, with $\mathcal{A}$ the adjacency matrix of the linear chain and $\mathcal{C}=(\mathcal{C}_1,...,\mathcal{C}_N)^t$. The diagonalization of the adjacency matrix provides the EGE, which is related to the particles distribution.\\
Alternatively, we can search for solutions of Eq. (\ref{eq:difflin}) compatible with the boundary conditions. Such solutions present the form $\mathcal{C}^{(m)}_n \propto \sin(k_m n)$ with $k_m=m \pi/(N+1)$. However, sign-changing solutions are not admissible and thus, after normalization, we get:
\begin{eqnarray}
\mathcal{C}^{(1)}_n =\sqrt{\frac{2 N_T}{N+1}} \sin(k_1 n),
\end{eqnarray}
with associated eigenvalue $\lambda=2\cos(k_1)$. $\mathcal{C}^{(1)}$ also corresponds to the ground state of the system with energy eigenvalue $\mu_0=-2K \cos(k_1)$. The fraction of particles belonging to the n-th superconducting island is given by the following relation:
\begin{eqnarray}
\label{eq:distrlin}
\frac{N_n}{N_T}=\frac{2}{N+1} \sin^{2}\Bigl(\frac{n \pi}{N+1}\Bigl),
\end{eqnarray}
showing the tendency to populate preferentially the central region of the system. This tendency originates from the fact that highly connected islands are preferentially occupied with respect to network's sites with connectivity deficiency (e.g., the system's boundaries). The aforementioned statement can be verified in simple cases since Eq. (\ref{eq:distrlin}) maintains its validity when linear chains with $N=1$, $N=2$ and $N=3$ network's islands are considered. Apart from the trivial case of a single network's site, we observe uniform occupation ($N_1=N_2=N_T/2$) when the case $N=2$ is considered, while, for $N=3$, inhomogeneous particles distribution with $N_1=N_3=N_T/4$ and $N_2=N_T/2$ is observed.

\begin{figure}[h]
\includegraphics[scale=0.45]{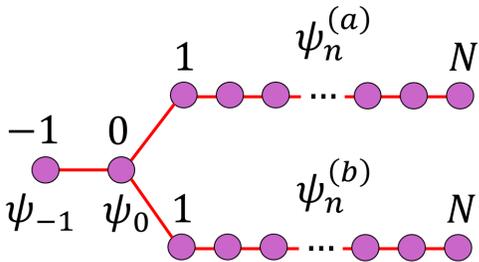}
\caption{Linear chain with a side-coupled island. The straight lines represent the connections between islands indicated by circles. The island labelled by $0$ presents the maximum degree ($\mathbb{D}_0=3$).}
\label{figure2}
\end{figure}

\subsection{Linear chain with a side-coupled island}
\label{sec:linside}
The wavefunction of $N_T$ Cooper pairs confined to move on a linear chain with a side-coupled site (Fig. \ref{figure2}) obeys the following equations:
\begin{eqnarray}
\label{eq:LCS1}
i \hbar \partial_t \psi_{-1} &=&-K \psi_0\\
\label{eq:LCS2}
i \hbar \partial_t \psi_{0} &=&-K (\psi_{-1}+\psi_{1}^{(a)}+\psi_{1}^{(b)})\\
\label{eq:LCS3}
i \hbar \partial_t \psi_{n}^{(\alpha)} &=& -K (\psi_{n-1}^{(\alpha)}+\psi_{n+1}^{(\alpha)})
\end{eqnarray}
where $n\geq 1$ is a labeling of the generic network site belonging to the branch $\alpha \in \{a,b\}$, while the additional relation $\psi_{0}^{(\alpha)}=\psi_0$ plays the role of a boundary condition. Using the exponential notation $\psi_{-1}=\sqrt{N_{-1}} \exp(i \theta)$, $\psi_{0}=\sqrt{N_{0}} \exp(i \theta)$ and $\psi_{n}^{(\alpha)}=\sqrt{N_{n}^{(\alpha)}} \exp(i \theta)$ in Equations (\ref{eq:LCS1})-(\ref{eq:LCS3}), we get
\begin{eqnarray}
\label{eq:lcs1}
\lambda A_{-1} &=& A_0\\
\label{eq:lcs2}
\lambda A_{0} &=& A_{-1}+A_1^{(a)}+A_1^{(b)}\\
\label{eq:lcs3}
\lambda A_{n}^{(\alpha)} &=& A_{n-1}^{(\alpha)}+A_{n+1}^{(\alpha)},
\end{eqnarray}
where we have introduced the notation $A_{-1}=\sqrt{N_{-1}}$, $A_{0}=\sqrt{N_{0}}$, $A_{n}^{(\alpha)}=\sqrt{N_{n}^{(\alpha)}}$ and $\lambda=\hbar \omega/K$. Numerical diagonalization of the problem shows that the EGE presents exponential localization. The latter statement can be easily proven by using the following ansatz:
\begin{eqnarray}
A_{-1} &=& \chi A_0\\
A_n^{(\alpha)} &=& A_0 \rho^n,
\end{eqnarray}
being $\chi$, $A_0$ and $\rho<1$ parameters to be determined. Using the ansatz in Eqs. (\ref{eq:lcs1})-(\ref{eq:lcs3}), we obtain the following relations:
\begin{eqnarray}
\label{eq:chi}
\chi &=& \lambda^{-1}\\
\label{eq:rho}
\rho &=& \frac{\lambda^2-1}{2\lambda}\\
\label{eq:lambda}
\lambda &=& \rho+\rho^{-1}.
\end{eqnarray}
Equations (\ref{eq:rho}) and (\ref{eq:lambda}) imply the equation $\lambda^4-4\lambda^2-1=0$, whose physically acceptable solution is given by $\lambda=\sqrt{2+\sqrt{5}}$. Once $\lambda$ is known, direct calculation shows that:
\begin{eqnarray}
\label{eq:ro}
\rho &=& \sqrt{\frac{\sqrt{5}-1}{2}} \approx 0.786\\
\chi &=& \frac{1}{\sqrt{2+\sqrt{5}}},
\end{eqnarray}
while $A_0$ is fixed by imposing the wavefunction normalization condition:
\begin{eqnarray}
A_{-1}^2+A_{0}^2+\sum_{\alpha \in \{a,b\}}\sum_{n=1}^{N}A_{n}^{(\alpha)2}=N_T.
\end{eqnarray}
Using the aforementioned results and assuming $N \gg 1$, the particles fraction allocated in the network sites labelled by $n=0$ and $n=-1$ is obtained:
\begin{eqnarray}
\frac{N_0}{N_T} &=& \frac{1}{2\sqrt{5}} \approx 0.2236\\
\frac{N_{-1}}{N_T} &=& \frac{1}{10+4\sqrt{5}} \approx 0.0527.
\end{eqnarray}
Moreover, the particles occupation along the $\alpha$ branch is easily obtained by using the relation $N_{n}^{(\alpha)}=N_0 \rho^{2n}$.\\
One of the main conclusions of this section is that particles manifest the tendency to localize exponentially around the system's island presenting the maximum connectivity. Exponential localization can be further characterized by introducing the localization length
\begin{eqnarray}
\label{eq:locL}
\mathcal{L}=-\frac{1}{2 \ln(\rho)},
\end{eqnarray}
being the latter solution of the equation $\rho^{2n}=\exp(-n/\mathcal{L})$. Using Eq. (\ref{eq:ro}) in Eq. (\ref{eq:locL}) one obtains $\mathcal{L} \approx 2.08$. An effective measure of the localization strength induced by the network connectivity can be obtained by comparing $\mathcal{L}$ with the localization length $\mathcal{L}_v$ induced by a local attractive potential, namely $-V$, acting on a specific island of a linear chain (see Appendix \ref{app:C}). The comparison is performed by solving the equation $\mathcal{L}=\mathcal{L}_v$ with respect to the variable $v$, being $v=V/K$. When this general procedure is applied to the present case, one obtains $v \approx 0.486$, implying a rather strong attractive potential $-0.486 \ K$. In this way, the localizing action of the side-coupled island has been measured in terms of a sort of effective potential.
\begin{figure}[h]
\includegraphics[scale=0.45]{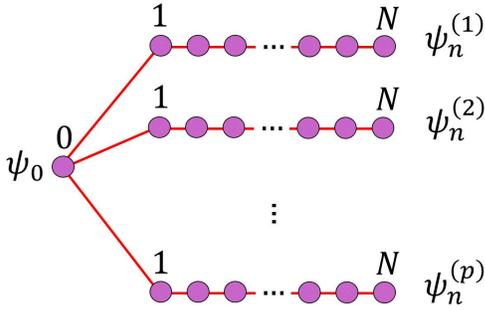}
\caption{Star-like network. The straight lines represent the connections between islands indicated by circles. The island labelled by $0$ presents the maximum degree ($\mathbb{D}_0=p$).}
\label{figure3}
\end{figure}

\subsection{Star-like network}
\label{subsec:star}
Star-like networks belong to a special class of tree-like structures (loop-free graphs) in which a central island is connected to several branches. In these systems, the degree of the central site is typically much higher than the degree of any other network's site and acts as localization center for particles. Apart from intrinsic motivations, star-like topology may describe parts of a more complex network and this observation further motivates the interest in studying these systems.\\
The wavefunction of $N_T$ Cooper pairs confined to move on a star-like network with $p$ branches (Fig. \ref{figure3}) obeys the following equations:
\begin{eqnarray}
\label{eq:star1}
i \hbar \partial_t \psi_{0} &=&-K \sum_{j=1}^{p}\psi_1^{(j)}\\
\label{eq:star2}
i \hbar \partial_t \psi_{n}^{(j)} &=&-K (\psi_{n-1}^{(j)}+\psi_{n+1}^{(j)})
\end{eqnarray}
with $n\geq 1$ and $j \in \{1,...,p\}$. Using exponential notation, i.e. $\psi_0=A_0 \exp(i \theta)$ and $\psi_n^{(j)}=A_n^{(j)} \exp(i \theta)$, along with the notation $A_0=\sqrt{N_0}$ and $A_n^{(j)}=\sqrt{N_n^{(j)}}$, we get the following relations:
\begin{eqnarray}
\label{eq:s1}
\lambda A_0 &=&\sum_{j=1}^{p} A_1^{(j)}\\
\label{eq:s2}
\lambda A_{n}^{(j)} &=& A_{n-1}^{(j)}+A_{n+1}^{(j)},
\end{eqnarray}
with the boundary condition $A_0^{(j)}=A_0$. Numerical diagonalization of the problem shows that the EGE presents exponential localization. The latter statement can be easily proven by using the ansatz $A_n^{(j)}=A_0 \rho^n$ with $\rho<1$ and $n\geq 1$. After the substitution in Equations (\ref{eq:s1}) and (\ref{eq:s2}), we obtain the relations
\begin{eqnarray}
\lambda &=& p \rho\\
\lambda &=& \rho+\rho^{-1},
\end{eqnarray}
which admit the following solutions:
\begin{eqnarray}
\lambda=\frac{p}{\sqrt{p-1}}\\
\rho=\frac{1}{\sqrt{p-1}}.
\end{eqnarray}
$A_0$ is fixed by imposing the normalization condition:
\begin{eqnarray}
A_0^{2}+\sum_{j=1}^{p} \sum_{n=1}^{N}A_n^{(j)2}=N_T.
\end{eqnarray}
Using the aforementioned results and assuming $N \gg 1$, the particles fraction allocated in the network site labelled by $n=0$ is obtained:
\begin{eqnarray}
\frac{N_0}{N_T} &=& \frac{1}{2}\Big(\frac{p-2}{p-1}\Big).
\end{eqnarray}
Interestingly, taking the limit $p\rightarrow\infty$ one observes $N_0 \rightarrow N_T/2$. Moreover, the particles occupation along the j-th branch is easily obtained by using the relation $N_n^{(j)}=N_0 \rho^{2n}$. The exponential localization of the particles distribution allows to use Eq. (\ref{eq:locL}) to extract the localization length in the form:
\begin{eqnarray}
\label{eq:locLStar}
\mathcal{L}=\frac{1}{\ln(p-1)},
\end{eqnarray}
while the parameter $v$ introduced at the end of Sec. \ref{sec:linside} takes the value:
\begin{eqnarray}
\label{eq:v}
v=\frac{p-2}{\sqrt{p-1}}.
\end{eqnarray}
When Equations (\ref{eq:locLStar}) and (\ref{eq:v}) are specialized to the case $p=3$, one obtains $\mathcal{L} \approx 1.44$ and $v \approx 0.707$. The latter estimates clearly show the relevance of connectivity-induced localization phenomena.\\
Cooper pairs wavefunction is expected to present the same profile exhibited by the superconducting order parameter. This expectation is indeed confirmed and we have found that the wavefunction of particles confined on a three-legs star-like network closely follows the superconducting order parameter profile derived in Ref. [\onlinecite{romeoGL}] in close vicinity of the critical temperature. Following Ref. [\onlinecite{romeoGL}], the order parameter profile deviates from a pure exponential when the temperature is lowered further. This deviation has no direct correspondence with the bosonic wavefunction since temperature is not included in the present treatment. Despite the latter observation, within the Ginzburg-Landau approach proposed in Ref. [\onlinecite{romeoGL}], a temperature lowering promotes a relevant influence of the nonlinear terms affecting the order parameter equations. As will be clearly explained in Sec. \ref{sec:starint}, a similar role is here played by the repulsive interaction, whose influence increases with the dimensionless parameter $\xi$, which is proportional to the bosons number $N_T$. Thus a full agreement between the distinct approaches implies that the parameter $N_T$ has to be interpreted as the average Cooper pairs number, being the latter a temperature-dependent quantity; it is vanishing above the normal-superconductor transition temperature, while it increases when the system is progressively cooled below the critical temperature.\\
According to Ref. [\onlinecite{romeoGL}], the order parameter localization characterizes the nucleation of the superconductivity in close vicinity of the critical temperature, while a spreading of the superconductivity to the whole system is observed when the temperature is lowered further. The mentioned behavior suggests that exponential localization of Cooper pairs characterizes the first insurgence of the superconductivity in star-like networks, while the initial localization is not able to drive the system towards a global insulating phase.

\begin{figure}[h]
\includegraphics[scale=0.35]{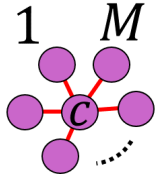}
\caption{Flower-like network. The straight lines represent the connections between islands indicated by circles. The island labelled by $c$ presents the maximum degree ($\mathbb{D}_c=M$).}
\label{figure4}
\end{figure}

\subsection{Flower-like network}
A flower-like network\cite{hubbardStar} (Fig. \ref{figure4}) is obtained by linking a central island with $M$ lateral sites, so that $N=M+1$ represents the total number of islands. Despite flower-like networks can be seen as a subcase of star-like structures, they deserve a separate treatment. Indeed, results presented in Sec. \ref{subsec:star} are here unapplicable because they are based on the assumption that the number of islands forming each branch of the star-like structure is much greater than one.\\
The wavefunction of particles constrained to a flower-like network is obtained by solving the following equations:
\begin{eqnarray}
\label{eq:flower1}
i\hbar \partial_t \psi_c &=& -K \sum_{j=1}^{M}\psi_j\\
\label{eq:flower2}
i\hbar \partial_t \psi_j &=& -K \psi_c,
\end{eqnarray}
where $\psi_c=\sqrt{N_c} \exp(i \theta_c)$ is related to the central island, while $\psi_j=\sqrt{N_j} \exp(i \theta_j)$ ($j \in \{1,..., M\}$) are related to the $M$ lateral sites. Using exponential representation, Equations (\ref{eq:flower1})-(\ref{eq:flower2}) take the following equivalent form:
\begin{eqnarray}
\dot{\theta}_c &=& \frac{K}{\hbar}\sum_{j=1}^{M}\sqrt{\frac{N_j}{N_c}} \cos(\theta_j-\theta_c)\\
\dot{N}_c &=& -\frac{2K}{\hbar}\sum_{j=1}^{M}\sqrt{N_c N_j} \sin(\theta_j-\theta_c)\\
\dot{\theta}_j &=& \frac{K}{\hbar} \sqrt{\frac{N_c}{N_j}} \cos(\theta_c-\theta_j)\\
\dot{N}_j &=& -\frac{2K}{\hbar}\sqrt{N_c N_j} \sin(\theta_c-\theta_j).
\end{eqnarray}
By imposing the stationary condition $\theta_c=\theta_j=\theta(t)$ with $\dot{\theta}(t)=\omega$, we get $\dot{N}_c=\dot{N}_j=0$ along with the additional relations:
\begin{eqnarray}
\lambda \sqrt{N_c} &=& \sum_{j=1}^{M} \sqrt{N_j}\\
\lambda \sqrt{N_j} &=& \sqrt{N_c}.
\end{eqnarray}
The above equations are easily solved so that, also considering the normalization condition, we get the islands' occupation:
\begin{eqnarray}
\label{eq:flocc1}
\frac{N_j}{N_T} &=& \frac{1}{2M}\\
\label{eq:flocc2}
\frac{N_c}{N_T} &=& \frac{1}{2}.
\end{eqnarray}
Moreover, $\lambda=\sqrt{M}$ so that the ground state energy is given by $\mu_0=-K \lambda$. Thus, we have demonstrated that the fraction of particles belonging to the central island does not depend on the number $M$ of lateral sites.

\section{Selected applications of the theory: interacting case ($U>0$)}
\label{sec:int}
The inclusion of repulsive interaction in the model produces in general non-trivial changes to the particles distribution on the network. However, there exist analytically solvable network models which can be used to validate the numerical procedure described in Sec. \ref{sec:NLSnum}. These models are presented in Sec. \ref{sec:flowerint} and \ref{sec:starlocint}. Once the iterative procedure has been validated, it is used to study the interacting star-like network model (Sec. \ref{sec:starint}), which is not analytically solvable.

\subsection{Interacting flower-like network}
\label{sec:flowerint}
Hereafter, we study a flower-like network obtained by linking a central island with $M$ lateral sites, so that $N=M+1$ represents the total number of islands. Under the assumption of non-vanishing repulsive interaction (i.e., $U>0$), the wavefunction of particles constrained to a flower-like network is obtained by solving the following equations:
\begin{eqnarray}
\label{eq:flowerInt1}
i\hbar \partial_t \psi_c &=& -K \sum_{j=1}^{M}\psi_j+U|\psi_c|^2 \psi_c\\
\label{eq:flowerInt2}
i\hbar \partial_t \psi_j &=& -K \psi_c+U|\psi_j|^2 \psi_j,
\end{eqnarray}
where $\psi_c$ is related to the central island, while $\psi_j$ ($j \in \{1,..., M\}$) are related to the $M$ lateral sites. Equations (\ref{eq:flowerInt1}) and (\ref{eq:flowerInt2}) can be solved by using the following ansatz:
\begin{eqnarray}
\psi_c=\sqrt{N_c}\exp(i \theta)\equiv A_c \exp(i \theta)\\
\psi_j=\sqrt{N_l}\exp(i \theta)\equiv A_l \exp(i \theta),
\end{eqnarray}
along with the normalization condition $N_c+M N_l=N_T$. Using the ansatz with the condition $\dot{\theta}=\omega$ in Equations (\ref{eq:flowerInt1}) and (\ref{eq:flowerInt2}) provides the following relations:
\begin{eqnarray}
\label{eq:nl1}
\lambda A_c &=& M A_l-u A_c^3\\
\label{eq:nl2}
\lambda A_l &=& A_c-u A_l^3
\end{eqnarray}
with $\lambda=\hbar \omega/K$ and $u=U/K$. The non-linear problem written in Eq. (\ref{eq:nl1})-(\ref{eq:nl2}) can be rewritten in terms of the auxiliary variable $z=A_c/A_l$ as follows:
\begin{eqnarray}
\label{eq:quartic}
z^4+\xi z(z^2-1)-M^2=0
\end{eqnarray}
with $\xi=u N_T$. Once the solution of the quartic equation in Eq. (\ref{eq:quartic}) is known, physical observables are obtained by using the following relations:
\begin{eqnarray}
\frac{N_c}{N_T} &=& \frac{z^2}{z^2+M}\\
\frac{N_l}{N_T} &=& \frac{1}{z^2+M}\\
\lambda &=& \frac{z^4-M}{z(z^2-1)}.
\end{eqnarray}
In particular, the non-interacting case (see Equations (\ref{eq:flocc1}) and (\ref{eq:flocc2})) is recovered by setting $\xi=0$ in Eq. (\ref{eq:quartic}) whose acceptable solution under this condition is given by $z=\sqrt{M}$.

\begin{figure}[h]
\includegraphics[scale=0.865]{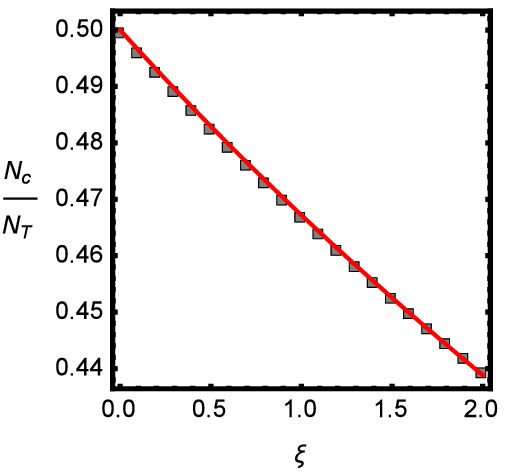}
\includegraphics[scale=0.8]{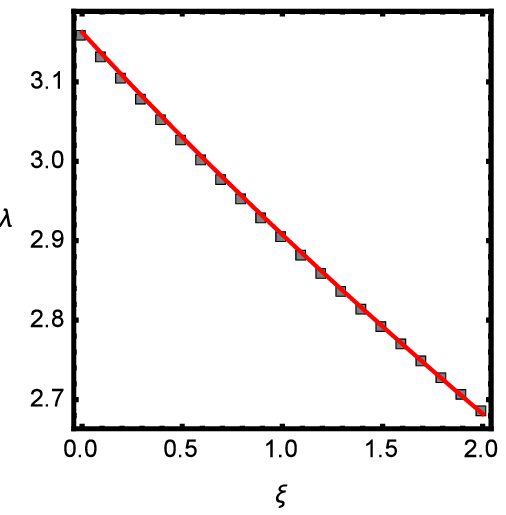}
\caption{Simulation of a flower-like network obtained by linking a central island with $M=10$ lateral sites, so that $N=11$ represents the total number of islands. Left panel: particle fraction $N_c/N_T$ allocated on the central site as a function of $\xi$. The full-line curve is well approximated by Eq. (\ref{eq:quarticApSol}) as long as $\xi\lesssim 0.5$. Right panel: Dependence of $\lambda$ on $\xi$. For both panels, square symbols are computed by using the iterative procedure described in Sec. \ref{sec:NLSnum}, while full line is obtained by using the approximate solution of Eq. (\ref{eq:quartic}) (see Appendix \ref{app:B} for details).}
\label{figure5ab}
\end{figure}

The model can be solved either by direct solution of Eq. (\ref{eq:quartic}) or by using the iterative procedure presented in Sec. \ref{sec:NLSnum}. The outcome of the model is presented in Fig. \ref{figure5ab}, where a flower-like network with $M=10$ lateral islands linked to a central site is studied. In particular, the particles fraction $N_c/N_T$ allocated on the central site as a function of $\xi$ is presented in left panel of Fig. \ref{figure5ab}. Repulsive interaction tends to reduce the  particles accumulation on a single site and for this reason the occupation of the central island is a decreasing function of $\xi$. Treating Eq. (\ref{eq:quartic}) as described in Appendix \ref{app:B} and considering the small-interaction limit ($\xi \ll 1)$, it is possible to demonstrate that the central island occupation is described by the following approximate relation:
\begin{eqnarray}
\label{eq:quarticApSol}
\frac{N_c}{N_T} \approx \frac{1}{2}-\frac{\xi(M-1)}{8 M^{3/2}}
\end{eqnarray}
as long as the interaction parameter $\xi$ is sufficiently small. Equation (\ref{eq:quarticApSol}) provides an excellent approximation of the curve shown in Fig. \ref{figure5ab} (left panel) capturing the linear decreasing of the central island occupation for $\xi \lesssim 0.5$. Considering the limit $M \gg 1$ in Eq. (\ref{eq:quarticApSol}), one observes that the competition between connectivity-induced localization and repulsive interaction is controlled by the parameter $\xi/\sqrt{M}$.\\
Moreover, increasing the repulsive interaction also produces a corresponding increasing of the ground state energy $\mu_0=-K \lambda$ (see right panel of Fig. \ref{figure5ab}).\\
We also observe that distinct solution methods show a good agreement and provide a consistent picture of the system's properties.

\subsection{Star-like network with repulsive interaction at the central island}
\label{sec:starlocint}
Hereafter, we focus on a star-like network in which the effect of the repulsive interaction between particles manifests only at the central island, while it is negligible elsewhere. This condition can be experimentally realized by considering a central island with reduced volume compared to that of the other network's islands. Thus, the wavefunction of $N_T$ Cooper pairs confined to move on a star-like network with $p$ branches and repulsive interaction at the central island obeys the following equations:
\begin{eqnarray}
\label{eq:staril1}
i \hbar \partial_t \psi_{0} &=&-K \sum_{j=1}^{p}\psi_1^{(j)}+U |\psi_0|^2 \psi_0\\
\label{eq:staril2}
i \hbar \partial_t \psi_{n}^{(j)} &=&-K (\psi_{n-1}^{(j)}+\psi_{n+1}^{(j)})
\end{eqnarray}
with $n\geq 1$ and $j \in \{1,...,p\}$. Using the notation $\psi_0=A_0 \exp(i \theta)\equiv \sqrt{N_0}\exp(i \theta)$ and $\psi_n^{(j)}=A_n^{(j)} \exp(i \theta)\equiv \sqrt{N_n^{(j)}} \exp(i \theta)$, we get the following relations:
\begin{eqnarray}
\label{eq:sil1}
\lambda A_0 &=&\sum_{j=1}^{p} A_1^{(j)}-u A_0^3\\
\label{eq:sil2}
\lambda A_{n}^{(j)} &=& A_{n-1}^{(j)}+A_{n+1}^{(j)},
\end{eqnarray}
with $u=U/K$ and the boundary condition $A_0^{(j)}=A_0$. Solution of the non-linear problem is obtained by using the ansatz $A_n^{(j)}=A_0 \rho^n$ with $\rho<1$ and $n\geq 1$. After the substitution in Equations (\ref{eq:sil1}) and (\ref{eq:sil2}), we obtain the following relations
\begin{eqnarray}
\lambda &=& p \rho-u A_0^2\\
\lambda &=& \rho+\rho^{-1}.
\end{eqnarray}
$A_0$ can be written in terms of $\rho$ by using the normalization condition. Accordingly, also assuming $N\gg 1$, we obtain:
\begin{eqnarray}
A_0^2=\frac{(1-\rho^2)N_T}{1+(p-1)\rho^2},
\end{eqnarray}
while $\rho$ obeys the quartic equation:
\begin{eqnarray}
\label{eq:quarticbis}
(p-1)^2 \rho^4+\xi \rho (\rho^2-1)-1=0,
\end{eqnarray}
with $\xi=u N_T$ and $\rho<1$. A close inspection of Eq. (\ref{eq:quarticbis}) clearly shows that repulsive interaction effects can be mitigated by the network topology (in graph theory sense). Indeed, by increasing the number of network's branches (i.e. for $p \gg 1$) the quantity $\xi/(p-1)^2$ becomes negligible and, under this condition, the system shows the tendency to behave as if the interaction at the central island were absent. The above observation corroborates the idea that highly connected islands are affected by topology-induced effective potentials and play the role of attractive centers of the particles' density.

Once the quartic equation has been solved, the occupation of the central island can be computed according to the relation:
\begin{eqnarray}
\frac{N_0}{N_T}=\frac{1-\rho^2}{1+(p-1)\rho^2},
\end{eqnarray}
while the occupation of islands belonging to a generic branch is given by $N_n^{(j)}=N_0 \rho^{2n}$.

\begin{figure}[h]
\includegraphics[scale=0.85]{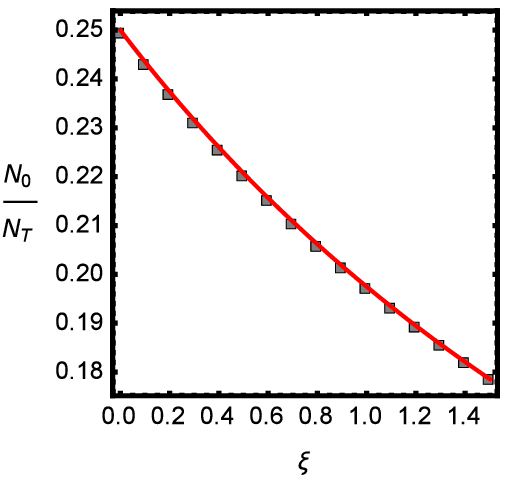}
\includegraphics[scale=0.8]{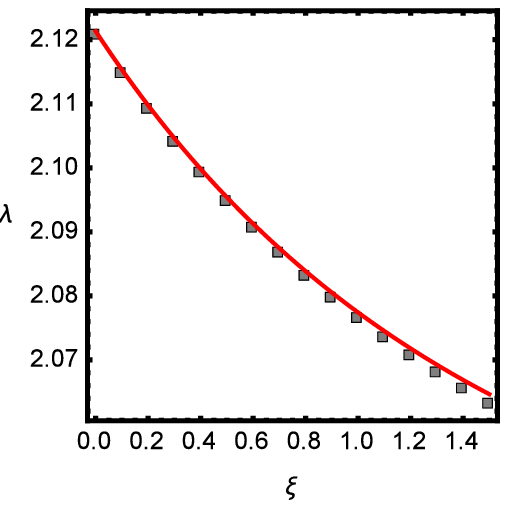}
\caption{Simulation of a star-like network with interaction localized on the central site. The network is obtained by linking a central site with three branches, each of which made of the connection of $25$ islands. Thus, $N=76$ represents the total number of islands.  Left panel: particles fraction $N_0/N_T$ allocated on the central site as a function of $\xi$. The full-line curve is well approximated by Eq. (\ref{eq:quartic2ApSol}) with $p=3$ as long as $\xi\lesssim 0.5$. Right panel: Dependence of $\lambda$ on $\xi$. For both panels, square symbols are computed by using the iterative procedure described in Sec. \ref{sec:NLSnum}, while full line is obtained by using the approximate solution of Eq. (\ref{eq:quarticbis}) (see Appendix \ref{app:B} for details).}
\label{figure6ab}
\end{figure}

The model can be solved either by direct solution of Eq. (\ref{eq:quarticbis}) or by using the iterative procedure presented in Sec. \ref{sec:NLSnum}. The outcome of the model is presented in Fig. \ref{figure6ab}, where a three-branches (i.e., $p=3$) star-like network with localized interaction is studied. In particular, the particles fraction $N_0/N_T$ allocated on the central site as a function of $\xi$ is presented in left panel of Fig. \ref{figure6ab}. As already observed before, the repulsive interaction tends to reduce the  particles accumulation on a single site and produces a corresponding increasing of the ground state energy $\mu_0=-K \lambda$ (see right panel of Fig. \ref{figure6ab}). Furthermore, the exponential ansatz used to derive Eq. (\ref{eq:quarticbis}) is validated by using the iterative procedure presented in Sec. \ref{sec:NLSnum}.\\
Using approximate methods described in Appendix \ref{app:B} and considering the small-interaction limit ($\xi\ll 1$), the central island occupation of a $p$-branch star-like network with localized interaction is described by the following relation:
\begin{eqnarray}
\label{eq:quartic2ApSol}
\frac{N_c}{N_T} \approx \frac{1}{2}\Big(\frac{p-2}{p-1}\Big)-\frac{\xi p(p-2)}{8 (p-1)^{5/2}}.
\end{eqnarray}
Equation (\ref{eq:quartic2ApSol}), with $p=3$, captures the linear decreasing of the central island occupation as a function of $\xi$ and provides an excellent approximation of the curve shown in Fig. \ref{figure6ab} (left panel) for $\xi \lesssim 0.5$. When Eq. (\ref{eq:quartic2ApSol}) with $p \gg 1$ is considered, one observes that the localization phenomena are controlled by the parameter $\xi/\sqrt{p}$.\\
Interestingly, Equations (\ref{eq:quarticApSol}) and (\ref{eq:quartic2ApSol}) take the same form as the limit $M\gg1$ and $p \gg 1 $, respectively, are considered. The latter observation suggests that, within the considered limit, the occupation of the central island is rather insensitive to the number of islands forming a single branch.

\subsection{Interacting Star-like network}
\label{sec:starint}
\begin{figure*}[t]
\includegraphics[scale=1.21]{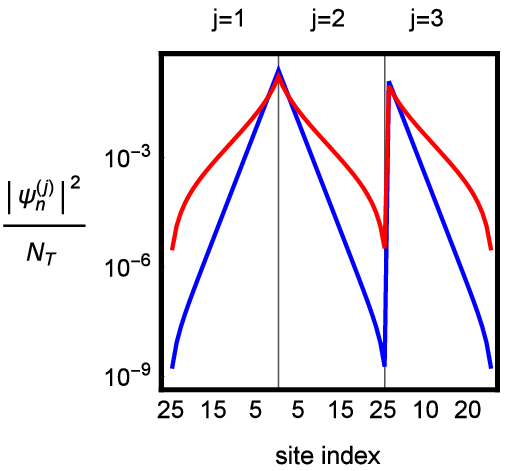} \
\includegraphics[scale=1.15]{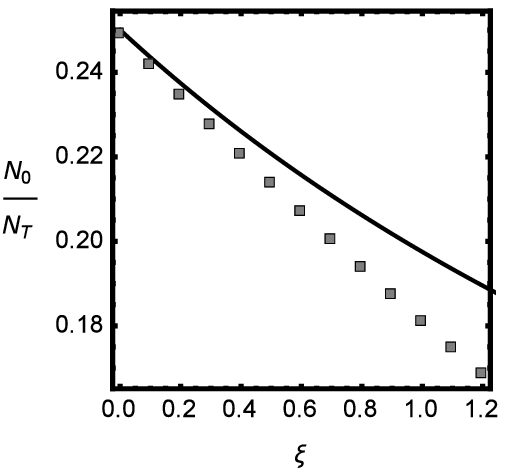} \
\includegraphics[scale=1.09]{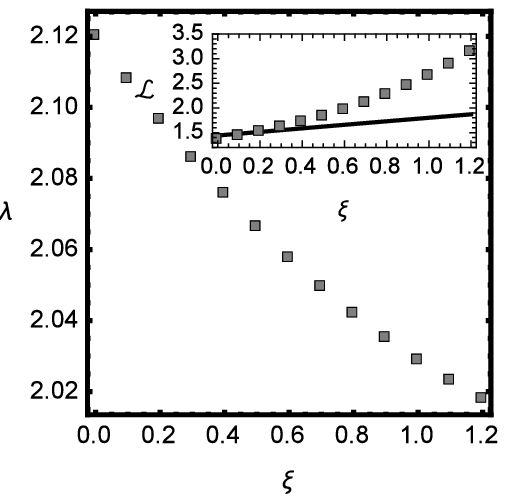}
\caption{Simulation of an interacting star-like network obtained by linking a central site with three branches, each of which made of the connection of $25$ islands. Thus, $N=76$ represents the total number of islands. Left panel: squared modulus of the wavefunction (red curve) computed by setting $\xi=1.2$ compared to the non-interacting (i.e., $\xi=0$) case (blue curve). The comparison, in logarithmic scale, evidences non-exponential behavior of the wavefunction describing the interacting system. Middle panel: particles fraction $N_0/N_T$ allocated on the central site as a function of $\xi$. Symbols are computed by using the iterative procedure described in Sec. \ref{sec:NLSnum}, while the full line represents the curve shown in Fig. \ref{figure6ab} (left panel), for comparison. Right panel: Dependence of $\lambda$ on $\xi$ obtained by using the procedure described in Sec. \ref{sec:NLSnum}. Inset: Localization length $\mathcal{L}$ versus $\xi$ of an interacting star-like network (symbols) and the corresponding behavior obtained for localized interaction (full line curve, see Sec. \ref{sec:starlocint} for details).}
\label{figure7abc}
\end{figure*}

The wavefunction of $N_T$ Cooper pairs confined to move on a star-like network with $p$ branches and repulsive interaction obeys the following equations:
\begin{eqnarray}
\label{eq:stari1}
i \hbar \partial_t \psi_{0} &=&-K \sum_{j=1}^{p}\psi_1^{(j)}+U |\psi_0|^2 \psi_0\\
\label{eq:stari2}
i \hbar \partial_t \psi_{n}^{(j)} &=&-K (\psi_{n-1}^{(j)}+\psi_{n+1}^{(j)})+U |\psi_{n}^{(j)}|^2 \psi_{n}^{(j)}
\end{eqnarray}
with $n\geq 1$ and $j \in \{1,...,p\}$. Within the same notation used in Sec. \ref{sec:starlocint}, we get the following relations:
\begin{eqnarray}
\label{eq:si1}
\lambda A_0 &=&\sum_{j=1}^{p} A_1^{(j)}-u A_0^3\\
\label{eq:si2}
\lambda A_{n}^{(j)} &=& A_{n-1}^{(j)}+A_{n+1}^{(j)}-u A_n^{(j)3},
\end{eqnarray}
with $u=U/K$ and the boundary condition $A_0^{(j)}=A_0$. Solution of the non-linear problem cannot be obtained by using the ansatz $A_n^{(j)}=A_0 \rho^n$ and thus we have to use the numerical procedure described in Sec. \ref{sec:NLSnum}.\\
In the following, we focus on the interacting star-like network obtained by linking a central site with three branches ($p=3$), each of which made of the connection of $25$ islands. Thus, $N=76$ represents the total number of islands. The particles distribution of the system with interaction parameter $\xi=1.2$ is shown in Figure \ref{figure7abc} (left panel). When the aforementioned distribution is compared with that pertaining to the non-interacting case (i.e. $\xi=0$), two distinctive features clearly appear. Indeed, interaction produces a non-exponential behavior of the particles distribution accompanied by a lowering of the particles fraction $N_0/N_T$ allocated at the central island. The latter tendency also populates islands belonging to the system's branches. This behavior is affected by the interaction strength. In particular, the particles fraction allocated at the central island is a decreasing function of the interaction parameter $\xi$ (see Figure \ref{figure7abc}, middle panel), which is an obvious signature of the repulsive interaction between particles. A comparison of the $N_0/N_T$ \textit{vs} $\xi$ curve with the corresponding shown in Fig. \ref{figure6ab} (left panel) evidences a similar behavior as long as the interaction parameter $\xi$ is smaller than $\sim 0.2$, while a marked deviation is observed for $\xi>0.2$. The latter observation suggests that, as long as $\xi\lesssim 0.1$, the $N_0/N_T$ \textit{vs} $\xi$ curve of an interacting star-like network is well described by Eq. (\ref{eq:quartic2ApSol}), which has been derived considering localized interaction (see Sec. \ref{sec:starlocint}). The presence of a finite repulsive interaction between particles is also responsible for an enhancement of the ground state energy of the system compared to the non-interacting case. This statement is easily verified by looking at the right panel of Figure \ref{figure7abc} where the dependence of $\lambda$ on $\xi$ is presented. Indeed, $\lambda$ is a decreasing function of the interaction strength $\xi$ and thus the ground state energy $\mu_0=-K \lambda$ increases due to the repulsive interaction.\\
The localization length as a function of the interaction parameter $\xi$ is reported in the inset of Fig. \ref{figure7abc} (right panel). The full-line curve represents the localization length $\mathcal{L}$ of a star-like network with interaction localized at the central island (see Sec. \ref{sec:starlocint}), while symbols represent the localization length of the interacting star-like network described in the present section. In the latter case, localization length has been extracted by considering the exponential decay of the particles distribution around the central part of the system branches. Localization length increases with $\xi$ evidencing the presence of interaction-driven delocalization phenomena. For $\xi>0.2$, the localization length of an interacting star-like network (symbols) grows faster as the interaction increases than the corresponding behavior observed in the case of localized interaction (full line curve). Differences are less relevant for small values of the interaction ($\xi<0.2$).\\
From the physics viewpoint, we observe that interacting star-like networks and star-like networks with interaction localized at the central island present a very similar behavior as long as the central island population includes a relevant fraction of the total number of particles (i.e., for small values of $\xi$). Under this condition, considering an interacting star-like network, islands belonging to the branches of the star are essentially depopulated and particles belonging to this part of the system behave as if the repulsive interaction were absent. Interaction always works at the central island. Intuition clearly indicates that the equivalence between the two models is broken as the interaction parameter $\xi$ increases. Indeed, under this condition, the central island occupation starts to be inhibited favoring the particles migration towards the system's branches. The increased number of particles belonging to the branches makes the repulsive interaction more effective.

\section{Discussion and conclusions}
\label{sec:concl}
We have presented a generalized Feynman's model to describe the inhomogeneous Cooper pairs distribution in tree-like networks of weakly-coupled superconducting islands. The theory allows the description of the interplay between particles interaction and network connectivity in graph-like artificial structures also evidencing interesting connections with the spectral graph theory. In these systems, the network connectivity defines effective potentials which are able to induce particles localization. Indeed, network's islands with higher connectivity act as localization centers, while the repulsive interaction works in the opposite way. Using a number of representative examples, it has been demonstrated that the particles fraction localized at a given system's site depends on the network's topology and, as expected, it is lowered by the repulsive interaction. In particular, we have demonstrated that the Cooper pairs fraction localized at the central site of a non-interacting flower-like network or at the center of a star-like network cannot exceed fifty percent of the total particles number $N_T$, irrespectively of any other detail. Moreover, the inclusion of repulsive interaction in star-like networks leads to depopulation of the central site and induces non-exponential localization phenomena.\\
Cooper pairs wavefunction is expected to present the same profile exhibited by the superconducting order parameter in view of the close connection between the aforementioned quantities. This expectation is indeed confirmed and we have found that the wavefunction of particles confined on a three-legs star-like network closely follows the superconducting order parameter profile derived in Ref. [\onlinecite{romeoGL}].\\
These findings suggest interesting analogies between disordered superconducting systems and tree-like networks of weakly-coupled superconducting islands. In both the mentioned systems localization phenomena are accompanied by an enhancement of the critical temperature of the superconducting transition, being the latter property appealing for possible technological applications.\\
Furthermore, since both Josephson arrays and optical lattices are experimental realizations of the Bose-Hubbard model, we expect that \textit{atomtronics} [\onlinecite{atomtronics}] platforms can be also used to test the localization phenomena reported in this work.

\section*{Acknowledgment}
Discussions with M. Cirillo, F. Corberi and M. Salerno are gratefully acknowledged.
\\
\appendix
\section{Properties of the ground state wavefunction}
\label{app:A}
Let us consider Eq. (\ref{eq:nse}) with $\epsilon_k=0$. Let be $\psi_k=|\psi_k|\exp(i \phi_k)\exp(i \theta(t))$ the wavefunction of a stationary state with energy eigenvalue $\mu=-\hbar \dot{\theta}=-\hbar \omega$. The stationary condition does not depend on the interaction and implies that:
\begin{eqnarray}
\label{eq:appst}
\frac{d}{dt}|\psi_{k}|^2=-\frac{2K}{\hbar} \sum_{j}\mathcal{A}_{kj}|\psi_k||\psi_j| \sin(\phi_j-\phi_k)=0.
\end{eqnarray}
Equation (\ref{eq:appst}) is respected when the phase difference between generic couples of adjacent sites $\phi_j-\phi_k$ is $0$ or $\pi$, implying that the Josephson bond currents are everywhere absent. This conclusion is easily reached by considering Eq. (\ref{eq:appst}) site by site, starting from nodes presenting a single connection to the rest of the network (nodes with degree one).  Moreover, a relation exists between the energy eigenvalue $\mu$ and the stationary state wavefunction. It can be written as:
\begin{eqnarray}
\label{eq:appmu}
\mu=-\frac{K}{N_{T}}\sum_{jk}\mathcal{A}_{kj}|\psi_k||\psi_j|e^{i(\phi_j-\phi_k)}+\sum_{k}\frac{U|\psi_k|^4}{N_T},
\end{eqnarray}
the second addendum being clearly affected by the interaction. According to Eq. (\ref{eq:appmu}), phase jumps of $\pi$ between adjacent sites increase the energy and thus it is expected that ground state wavefunction can be written in the form $\psi_k=|\psi_k| \exp(i\phi) \exp(i \theta(t))$, being $\exp(i\phi)$ a global (time-independent) phase factor. Omitting the irrelevant factor $\exp(i\phi)$, the ground state wavefunction can be presented in the form $\psi_k=|\psi_k|\exp(i \theta(t))$ with $|\psi_k|=\sqrt{N_k}$.

\section{Confining properties of a local attractive potential}
\label{app:C}
\begin{figure}[!h]
\includegraphics[scale=0.45]{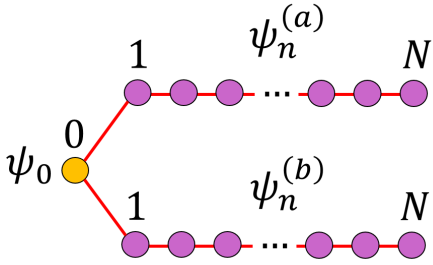}
\caption{Linear chain with a local potential on the central island. The straight lines represent the connections between islands indicated by circles. The island labelled by $0$ is affected by the local potential.}
\label{figure8}
\end{figure}

Let us consider a linear chain of islands presenting a local attractive potential ($-V$) on a central island labeled by $0$ (Fig. \ref{figure8}). The time evolution of the bosonic wavefunction is described by the following equations:
\begin{eqnarray}
i\hbar \partial_t \psi_0 &=& -V \psi_0-K (\psi^{(a)}_1+\psi^{(b)}_1)\\
i\hbar \partial_t \psi^{(\alpha)}_n &=& -K (\psi^{(\alpha)}_{n+1}+\psi^{(\alpha)}_{n-1}),
\end{eqnarray}
where $n\geq1$ is a labeling of the generic network site belonging to the branch $\alpha \in \{a,b\}$, while the additional relations $\psi^{(\alpha)}_0=\psi_0$ play the role of boundary conditions. Adopting a solution method similar to the one used in Sec. \ref{sec:linside} of the main text, we obtain that particles are distributed according to the relation $N^{(\alpha)}_n=N_0 \rho^{2n}$, being $N_0$ the occupation of the central island and
\begin{eqnarray}
\label{eq:Appc1}
\rho=\frac{\sqrt{v^2+4}-v}{2},
\end{eqnarray}
with $v=V/K>0$. The confining properties of the local potential can be characterized by introducing the localization length $\mathcal{L}_{v}$ according to the relation:
\begin{eqnarray}
\label{eq:Appc2}
\mathcal{L}_{v}=-\frac{1}{2 \ln (\rho)},
\end{eqnarray}
being the latter solution of the equation $\exp(-n/\mathcal{L}_v)=\rho^{2n}$. A closer inspection to Equations (\ref{eq:Appc1}) and (\ref{eq:Appc2}) shows that $\mathcal{L}_{v}$ is a decreasing function of $v$. Moreover, particles localization increases when $\mathcal{L}_{v}$ decreases.

\section{Perturbative solution of Eq. (\ref{eq:quartic}) and (\ref{eq:quarticbis})}
\label{app:B}
Despite $4$th order algebraic equations admit exact solutions (Ferrari-Cardano method), it is here useful searching for approximate roots of Eq. (\ref{eq:quartic}) and (\ref{eq:quarticbis}). Approximate roots are expressed via relatively simple mathematical expressions and can be used to understand the interplay between topological confinement effects and repulsive interaction. Both the aforementioned equations can be written in the general form $f(s)=0$ with
\begin{eqnarray}
f(s)=s^4+\alpha s(s^2-1)-\beta
\end{eqnarray}
and $\beta>0$. Moreover, the parameter $\alpha$ is proportional to the dimensionless interaction strength $\xi$. We are interested in finding a real and positive solution. If $\alpha$ is sufficiently small, the exact solution exhibits a small deviation from the solution of the non-interacting problem $s_0=\sqrt[4]{\beta}$. Let us call $\delta s\ll s_0$ the aforementioned deviation and write the equation $f(s_0+\delta s)=0$. Disregarding cubic and quartic terms in $\delta s$, we obtain the following equation:
\begin{eqnarray}
(6 s_0^2+3 \alpha s_0) \delta s^2+(4 s_0^3+3 \alpha s_0^3-\alpha) \delta s+ \alpha s_0(s_0^2-1)=0.\nonumber
\end{eqnarray}
The admissible solution of previous equation is given by:
\begin{widetext}
\begin{eqnarray}
\delta s=\frac{-4 s_0^3+(1-3s_0^2)\alpha+\sqrt{16 s_0^6+\alpha^2+\alpha s_0^2(16 s_0+6 \alpha-3 \alpha s_0^2)}}{6 s_0 (2 s_0+\alpha)}.
\end{eqnarray}
\end{widetext}
Let us note that $\delta s\rightarrow 0$ for vanishing interaction (i.e., for $\alpha \rightarrow 0$). Thus, the approximate solution of the equation $f(s)=0$ is given by $s_0+\delta s$.


\begin{thebibliography}{99}
\bibitem{nota1}
In graph theory, the degree $\mathbb{D}_i$ of a node indexed by $i$ in an undirected graph represents the number of links connecting it with other nodes in the network. The degree of a node can be expressed in terms of the adjacency matrix elements as $\mathbb{D}_i=\sum_j \mathcal{A}_{ij}$. The maximum degree of a graph is given by the number max$\{\mathbb{D}_1,...,\mathbb{D}_N \}$, being the latter computed considering the degree $\mathbb{D}_j$ of each network's node indexed by $j \in\{1,..., N\}$.
\end{thebibliography}
\end{document}